\documentclass[letter,11pt]{article}
\pdfoutput=1
 \usepackage{jheppub}
 
 \usepackage{graphicx}
 \usepackage{hyperref}
 \usepackage[utf8]{inputenc}
 \usepackage{amssymb}
\usepackage{multirow}
\usepackage{soul}
 
 \usepackage{booktabs}
\usepackage{mathrsfs}
\usepackage{epsfig}
\usepackage{graphicx}
\usepackage{subfigure}
\usepackage{dcolumn}
\usepackage{bm}
\usepackage{amsmath}
\usepackage{slashed}
\usepackage{multirow}
\usepackage{color}
\usepackage{dcolumn}
\usepackage{bm}
\usepackage{color}

\allowdisplaybreaks

 \newcommand{\lsim}{{\;\raise0.3ex\hbox{$<$\kern-0.75em\raise-1.1ex\hbox{$\sim$}}\;}}
\newcommand{\gsim}{{\;\raise0.3ex\hbox{$>$\kern-0.75em\raise-1.1ex\hbox{$\sim$}}\;}}
\newcommand{\beq}{\begin{equation}}
\newcommand{\eeq}{\end{equation}}
\newcommand{\bea}{\begin{eqnarray}}
\newcommand{\eea}{\end{eqnarray}}

\def\baa{\begin{array}}
\def\eaa{\end{array}}

\mathchardef\minus="002D

\preprint{ }

\title{Type-II Seesaw Leptogenesis along the Ridge}

\author{Chengcheng Han$^{1,2}$,}
\emailAdd{hanchch@mail.sysu.edu.cn}

\author{Zhanhong Lei$^{1}$, }
\emailAdd{leizhh3@mail2.sysu.edu.cn}

\author{Jin Min Yang$^{3,4}$}
\emailAdd{jmyang@itp.ac.cn}

\vspace{0.5cm}

\affiliation{$^1$School of Physics, Sun Yat-Sen University, Guangzhou 510275, P. R. China}
\affiliation{$^2$Asia Pacific Center for Theoretical Physics, Pohang 37673, Korea}
\affiliation{$^3$CAS Key Laboratory of Theoretical Physics, Institute of Theoretical Physics, Chinese Academy of Sciences, Beijing 100190, P. R. China}
\affiliation{$^4$Department of Physics, Henan Normal University, Xinxiang 453007,  P. R. China}

\vspace{0.5cm}

\abstract{ Type-II seesaw leptogenesis is a model that integrates inflation, baryon number asymmetry, and neutrino mass simultaneously. It employs the Affleck-Dine mechanism to generate lepton asymmetry, with the Higgs bosons serving as the inflaton. Previous studies assumed inflation to occur in a valley of the potential, employing the single-field approximation. In this work, we explore an alternative scenario for the type-II seesaw leptogenesis, where the inflation takes place along a ridge of the potential. Firstly, we conduct a comprehensive numerical calculation in the canonical scenario, where inflation occurs in a valley, confirming the effectiveness of the single-field approximation. Then, we introduce a novel scenario wherein inflation initiates along the potential's ridge and transitions to the valley in the late stages. In this case, the single-field inflation approximation is no longer valid, yet leptogenesis is still successfully achieved. We find that this scenario can generate a significant non-Gaussianity signature, offering testable predictions for future experiments.
}

\makeatletter
\def\@fpheader{\relax}
\makeatother

\date{\today}

\begin{document} 
\maketitle
\flushbottom
\newpage

\section{Introduction}
Many open problems remain to be solved in particle physics and cosmology, including the three typical ones: the origin of neutrino masses, the matter-antimatter asymmetry, and the nature of inflation.  To solve these problems, various models have been proposed, among which the seesaw mechanisms~\cite{Minkowski:1977sc, Yanagida:1979as, Glashow:1979nm, GellMann:1980vs, Magg:1980ut,Cheng:1980qt,Lazarides:1980nt,Mohapatra:1980yp, Foot:1988aq,Albright:2003xb} are rather popular since they can interpret the tiny neutrino masses and the baryon asymmetry through the leptogenesis~\cite{Fukugita:1986hr}. There are also attempts to simultaneously solve these three problems, usually through considering some extensions to the seesaw models~\cite{Murayama:1992ua, Murayama:1993xu, Senami:2001qn, Evans:2015mta, Mohapatra:2021ozu, Mohapatra:2022tgb, Hertzberg:2013mba, Lozanov:2014zfa, Yamada:2015xyr, Bamba:2016vjs,Bamba:2018bwl, Cline:2019fxx,Barrie:2020hiu, Lin:2020lmr, Kawasaki:2020xyf, Kusenko:2014lra, Wu:2019ohx, Charng:2008ke, Ferreira:2017ynu, Babichev:2018sia, Rodrigues:2020dod, Lee:2020yaj, Enomoto:2020lpf, Lloyd-Stubbs:2020sed, Mohapatra:2021aig, Mohapatra:2022ngo}. 

Interestingly, recent studies~\cite{Barrie:2021mwi,Barrie:2022cub} have shown that the minimal Type-II seesaw may address the above three problems simultaneously. In this framework, the inflaton is provided by the mixing of the SM Higgs $H$ with the triplet scalar $\Delta$ in the Type-II seesaw model. The nonminimal coupling of these two scalars with gravity induces a flat direction, along which the slow-roll inflation can be achieved. Considering the Affleck-Dine mechanism~\cite{Affleck:1984fy}, as the inflaton evolves along the inflationary trajectory, the phase of $\Delta$ carrying lepton number also evolves to give rise to a net lepton number density. Then, after the inflation ends and the universe thermalizes, the asymmetry of lepton number is converted into the asymmetry of baryon number through the sphaleron process~\cite{Kuzmin:1985mm}.  Finally, after spontaneous electroweak symmetry breaking, the neutral component of $\Delta$ obtains a nonzero vev, which is responsible for the neutrino masses. 

In \cite{Barrie:2021mwi,Barrie:2022cub} the authors considered the large field approximation and the inflation can be effectively described as a single field. In this work we conduct the calculation numerically and show that under certain conditions, the inflaton indeed evolve along a single flat direction, which is the valley of the potential. Even if the initial values of the fields are chosen arbitrarily, the inflaton will fall quickly into the valley and the subsequent evolution is essentially analogous to single-field  slow-roll inflation, known as the attractor solution. In addition, we observe that there exist other cases where the evolution of inflaton deviates from the standard single-field  slow-roll inflation. Particularly, if there exist a local minimum of the potential at the top of the ridge, the inflaton initially on the ridge will evolve along the ridge for a period, and then roll off and evolve along the valley until the inflation ends. Our calculations show that as long as the inflaton rolls off the ridge late enough, then the produced net lepton number density  will not be diluted too much by the subsequent inflation and the baryon asymmetry of the universe can be interpreted. Moreover, the deviation from the single-field  approximation may give rise to large primordial non-Gaussianity, which is an important feature to distinguish inflation models. Using the $\delta N$ formalism, we calculate the non-Gaussianity and find that the successful Type-II seesaw leptogenesis along the ridge can give rise to a sizable $|f_{NL}|$, which can be tested by the upcoming CMB measurements. Other phenomenology studies on the type-II seesaw leptogenesis can be found in~\cite{Han:2023vme, Han:2022ssz, Barrie:2022ake}.

This paper is organized as follows. In Sec.~\ref{sec:typeIIseesaw} we briefly review the inflation dynamics under the Type-II seesaw framework. In Sec. \ref{sec:leptogenesis} we analyse the Affleck-Dine mechanism along the inflationary trajectory. We conduct the calculation in the cases where the inflaton evolves along the valley and along the ridge respectively, and show that successful leptogenesis can be achieved in both cases. After that, we calculate the non-Gaussianity produced in the second case in Sec.~\ref{sec:nongauss}. Finally, in Sec.~\ref{sec:conclusion} we give a summary and draw conclusions. 

\section{Type-II Seesaw and Inflation}
\label{sec:typeIIseesaw}
In this section, we briefly overview the inflation dynamics in type-II seesaw model. The nonminimal couplings of the triplet Higgs in type-II seesaw model and the SM Higgs to gravity induces a flat potential for large fields and induce a Starobinsky type inflation in the early universe~\cite{Starobinsky:1980te,Bezrukov:2007ep,Bezrukov:2008ut,GarciaBellido:2008ab,Barbon:2009ya,Barvinsky:2009fy,Bezrukov:2009db,Giudice:2010ka,Bezrukov:2010jz,Burgess:2010zq,Lebedev:2011aq, Hamada:2014iga, Lee:2018esk,Choi:2019osi}. The multifield inflation dynamics is analysed in the Einstein frame, under which the inflation trajectory can be deduced from the equation of motion of the inflaton.

\subsection{Type-II seesaw}
The type-II seesaw mechanism attempts to explain neutrino masses through introducing an $SU(2)_L$ triplet scalar $\Delta$ to the SM, which carries a hypercharge $Y=1$. The triplet and SM Higgs are parameterized as 
\begin{equation}
H=\left(
\begin{array}{cc}
     h^+  \\
     h
\end{array}
\right),~~
\Delta=\left(
\begin{array}{cc}
    \Delta^+/\sqrt{2} & \Delta^{++} \\
    \Delta^0 & -\Delta^+/\sqrt{2}
\end{array}
\right),
\end{equation}
where $h$ and $\Delta^0$ are the neutral components. The triplet $\Delta$ induces a gauge invariant, renormalizable Yukawa term,
\begin{equation}
    \mathcal{L}_{\text {Yukawa }}=\mathcal{L}_{\text {Yukawa }}^{\mathrm{SM}}-\frac{1}{2} y_{j k} \overline{L}_{j}^{c} i\sigma^2 \Delta L_{k}+\text { h.c. },
\end{equation}
where $L_i$ represents a SM left-handed lepton doublet. Thus we can assign a lepton charge of $Q_L=-2$ to $\Delta$. After electroweak symmetry breaking, this term generates a neutrino mass matrix $m_{i j}^{\nu}=y_{i j} v_{\Delta}$ when $\Delta^0$ obtains a non-zero vev. Diagonalizing $m^\nu_{i j}$ by the PMNS matrix gives the Majorana neutrino masses. 

The field $\Delta$ also induces new terms to the scalar potential, given by
\begin{eqnarray}
\label{eq:potential}
    V(H,\Delta)&=&-m^2_HH^\dagger H+m_\Delta^2\text{Tr}(\Delta^\dagger\Delta) +\lambda_H(H^\dagger H)^2+\lambda_1(H^\dagger H)\text{Tr}(\Delta^\dagger\Delta) \nonumber \\
    &&+\lambda_2\left(\text{Tr}(\Delta^\dagger\Delta)\right)^2 +\lambda_3\text{Tr}(\Delta^\dagger\Delta)^2+\lambda_4H^\dagger\Delta\Delta^\dagger H +\left[\mu(H^Ti\sigma^2\Delta^\dagger H) \right.\nonumber \\
   && \left. +\frac{\lambda_5}{M_P}(H^Ti\sigma^2\Delta^\dagger H)(H^\dagger H)+ \frac{\lambda_5^\prime}{M_P}(H^Ti\sigma^2\Delta^\dagger H)(\Delta^\dagger \Delta) +h.c.\right]~.
\end{eqnarray}
Here the terms in the bracket lead to lepton number violation, which is necessary for successful leptogenesis. Two dimension-5 operators are also included, which are suppressed at low energy but would be important during the early universe. 


\subsection{Inflation from Higgs}

The CMB observation indicates that the inflaton should evolve along a flat potential. As is known in the scenario of Higgs inflation, the nonminimal coupling of Higgs scalars to the Ricci scalar $R$  induces a flat direction in the large field limit. In our framework, the Lagrangian in the Jordan frame with nonminimal coupling is given by
\begin{eqnarray}
    \label{con:lagrangian}
    \begin{aligned}
    \frac{\mathcal{L}}{\sqrt{-g}}=&\frac{1}{2}M^2_PR+\xi_HH^\dagger HR+\xi_\Delta \text{Tr}(\Delta^\dagger \Delta)R\\
    &-g^{\mu\nu}(D_\mu H)^\dagger(D_\nu H)-g^{\mu\nu}\text{Tr}(D_\mu \Delta)^\dagger(D_\nu \Delta)-V(H,\Delta)+\mathcal{L}_{Yukawa}~.
    \end{aligned}
\end{eqnarray}

The following analysis will focus on the neutral components of the scalars which have nonzero vev, thus the cosmological relevant Lagrangian is given by
\begin{eqnarray}
\label{lag:jordan}
\begin{aligned}
    \frac{\mathcal{L}}{\sqrt{-g}}=&\frac{1}{2}M^2_PR+\xi_H|h|^2R+\xi_\Delta |\Delta^0|^2R-(\partial_\mu h)^2-(\partial_\mu \Delta^0)^2-V(h,\Delta^0)\\
    =&\frac{1}{2}(M_P^2+\xi_H\rho_H^2+\xi_\Delta\rho_\Delta^2)R-\frac{1}{2}(\partial_\mu \rho_H)^2-\frac{1}{2}(\partial_\mu \rho_\Delta)^2-V(h,\Delta^0)
\end{aligned}~~,
\end{eqnarray}
and the scalar potential can be simplified as 
\begin{eqnarray}
    V(h,\Delta^0)&=&-m_H^2|h|^2+m_\Delta^2|\Delta^0|^2+\lambda_H|h|^4+\lambda_\Delta |\Delta^0|^4+\lambda_{H\Delta}|h|^2|\Delta^0|^2\nonumber \\
   &&  -\left(\mu h^2 {\Delta^0}^* + \frac{\lambda_5}{M_P} |h|^2  h^2  {\Delta^0}^*  + \frac{\lambda^\prime_5}{M_P} |\Delta^0|^2 h^2  {\Delta^0}^* +h.c. \right) +...  ~,
\end{eqnarray}
where $\lambda_\Delta=\lambda_2+\lambda_3$ and $\lambda_{H\Delta}=\lambda_1+\lambda_4$. We have introduced the polar coordinate parameterizations
\begin{equation}
    h\equiv\frac{1}{\sqrt{2}}\rho_He^{i\eta},~~~~\Delta^0\equiv\frac{1}{\sqrt{2}}\rho_\Delta e^{i\theta}.
\end{equation}
Since the potential is dominated by the radial directions $\rho_H$ and $\rho_\Delta$, the angular motion of $\eta$ and $\theta$ can be ignored in the analysis of inflationary dynamics. 

After a Weyl transformation,
\begin{equation}
    \label{con:transform}
    g_{E\mu\nu}=\Omega^2 g_{\mu\nu},~~\Omega^2=1+\frac{\xi_H\rho_H^2}{M_P^2}+\frac{\xi_\Delta\rho_\Delta^2}{M_P^2}~,
\end{equation}
we obtain the Lagrangian in the Einstein frame,
\begin{equation}
    \label{lag:einstein}
    \frac{\mathcal{L}}{\sqrt{-g_E}}=\frac{1}{2}M_P^2R_E-\frac{1}{2\Omega^2}\left((\partial_\mu \rho_H)^2+(\partial_\mu \rho_\Delta)^2\right)-3M_P^2(\partial_\mu \text{log}\Omega)^2-\frac{V(h,\Delta^0)}{\Omega^4}~.
\end{equation}

In the following we analyse the inflation dynamics by considering the action of multifield inflation. We denote $\phi^I=(\rho_H,\rho_\Delta)$, then from Eq. \ref{lag:einstein}, the action in the Einstein frame is 
\begin{equation}
\label{eq:action_multifield}
	S=\int d^{4} x \sqrt{-g_E} \times\left[\frac{M_{P}^{2}}{2} R-\frac{1}{2} G_{I J} g_E^{\mu \nu} \partial_{\mu} \phi^{I} \partial_{\nu} \phi^{J}-V_E\left(\phi\right)\right]~,
\end{equation}
where $V_E(\phi)=V(h,\Delta^0)/\Omega^4$ and $G_{I J}$ is the field metric given by
\begin{equation}
    G_{I J}=\frac{M_{P}^{2}}{2 f\left(\phi^{I}\right)}\left[\delta_{IJ}+\frac{3}{f\left(\phi^{I}\right)} f_{, I} f_{, J}\right]~,
\end{equation}
with $f(\phi^I)=\frac{1}{2}(M^2_P+\xi_H\rho^2_H+\xi_\Delta\rho^2_\Delta)$ being the nonminimal coupling function and $f_{,I}=\partial f/\partial \phi^I$.

Varying the inflation action with respect to $g_{E\mu\nu}$ and $\phi^I$, and focusing on the background part of inflaton, $\phi^I =\varphi^I + \delta\phi^I$, we obtain the Friedmann equations and the Klein-Gordon equation,
\begin{equation}
	\begin{aligned}
	H^{2}&=\frac{1}{3 M_{P}^{2}}\left[\frac{1}{2} G_{I J} \dot{\varphi}^{I} \dot{\varphi}^{J}+V\right]~, \\
	\dot{H}&=-\frac{1}{2 M_{P}^{2}} G_{I J} \dot{\varphi}^{I} \dot{\varphi}^{J}~,
	\end{aligned}
\end{equation}
\begin{equation}
\label{eq:back_evolution}
	\mathcal{D}_t\dot{\varphi}^{I}+3 H \dot{\varphi}^{I}+G^{I K} V_{, K}=0~,
\end{equation}
where $\mathcal{D}_{t} A^{I} \equiv \dot{\varphi}^{J} \mathcal{D}_{J} A^{I}$, and the covariant derivative in the field space is $\mathcal{D}_{J} A^{I}=\partial_{J} A^{I}+\Gamma_{J K}^{I} A^{K}$. By numerically solving Eq.~\ref{eq:back_evolution}, we can obtain the evolution of the background trajectory. 

In the case of multifield inflation, the evolution of the inflaton may be complicated, which, however, can be drastically simplified in the so-called kinematical basis~\cite{Gordon:2000hv,Peterson:2010np,Peterson:2011yt}. We can define the speed of the field vector as
\begin{equation}
	\dot{\sigma}=\left|\dot{\varphi}^{I}\right|=\sqrt{G_{I J} \dot{\varphi}^{I} \dot{\varphi}^{J}}~,
\end{equation} 
then the Friedmann equations simplify to
\begin{equation}
	H^{2}=\frac{1}{3 M_{P}^{2}}\left[\frac{1}{2} \dot{\sigma}^{2}+V\right], \quad \dot{H}=-\frac{1}{2 M_{P}^{2}} \dot{\sigma}^{2}~.
\end{equation}
It can be seen that $\dot{\sigma}$ captures the evolution characteristics along the field trajectory, which is similar to the single-field case. Thus the slow-roll parameters can be defined as in single-field inflation,
\begin{equation}
	\begin{array}{l}
	\epsilon \equiv-\frac{\dot{H}}{H^{2}}=\frac{ 3\dot{\sigma}^{2}}{\left(\dot{\sigma}^{2}+2 V\right)}~,\\
	\eta_{\sigma \sigma} \equiv M_{P}^{2} \frac{\mathcal{M}_{\sigma \sigma}}{V}~,
	\end{array}
\end{equation}
where $\mathcal{M}_{\sigma \sigma} =\hat{\sigma}^{K} \hat{\sigma}^{J}\left(\mathcal{D}_{K} \mathcal{D}_{J} V\right)$, and $\hat{\sigma}^I=\dot{\varphi}^I/\dot{\sigma}$. 

The deviation from single-field inflation is captured by the turn rate $\omega^I=\mathcal{D}_t \hat{\sigma}^I$, which represents how quickly the field trajectory is changing direction. If the so-called `slow-turn limit' $|\omega^I|\ll 1$ is satisfied, then the inflation dynamics is effectively single-field inflation.

\subsection{The potential and inflation trajectory }
\label{sec:potential}
In the following we adopt the natural unit $M_P=1$ with $M_P$ being the Planck scale. Considering $\xi_H\rho_H^2+\xi_\Delta\rho_\Delta^2\gg 1$ during inflation, the kinetic Lagrangian of the scalar fields is
\begin{equation}
    \label{con:kinetic}
    \mathcal{L}_{kin}=-\frac{3}{4}\left(\partial_\mu \text{log}(\xi_H\rho_H^2+\xi_\Delta\rho_\Delta^2)\right)^2-\frac{1}{2(\xi_H\rho_H^2+\xi_\Delta\rho_\Delta^2)}\left((\partial_\mu \rho_H)^2+(\partial_\mu \rho_\Delta)^2\right)~,
\end{equation}
which is noncanonical. Thus we redefine the fields as
\begin{equation}
    \label{con:redefine}
    \chi=\sqrt{\frac{3}{2}}\text{log}(\xi_H\rho_H^2+\xi_\Delta\rho_\Delta^2),~~\kappa=\frac{\rho_H}{\rho_\Delta}~.
\end{equation}
For large nonminimal couplings $\xi_H,~\xi_\Delta\gg 1$, the mixing term $(\partial_\mu\chi)(\partial^\mu\kappa)$ is suppressed, and the kinetic Lagrangian reduces to 
\begin{equation}
    \mathcal{L}_{kin}=-\frac{1}{2}(\partial_\mu\chi)^2-\frac{1}{2}\frac{\xi_H^2\kappa^2+\xi_\Delta^2}{(\xi_H\kappa^2+\xi_\Delta)^3}(\partial_\mu\kappa)^2~,
\end{equation}
in which $\chi$ is canonically normalised. A canonically normalized field $\kappa^\prime$ can also be defined in different regimes for $\kappa$. 

\begin{figure}[htbp]
    \centering
    \subfigure[case(1)]{\includegraphics[width=0.48\textwidth]{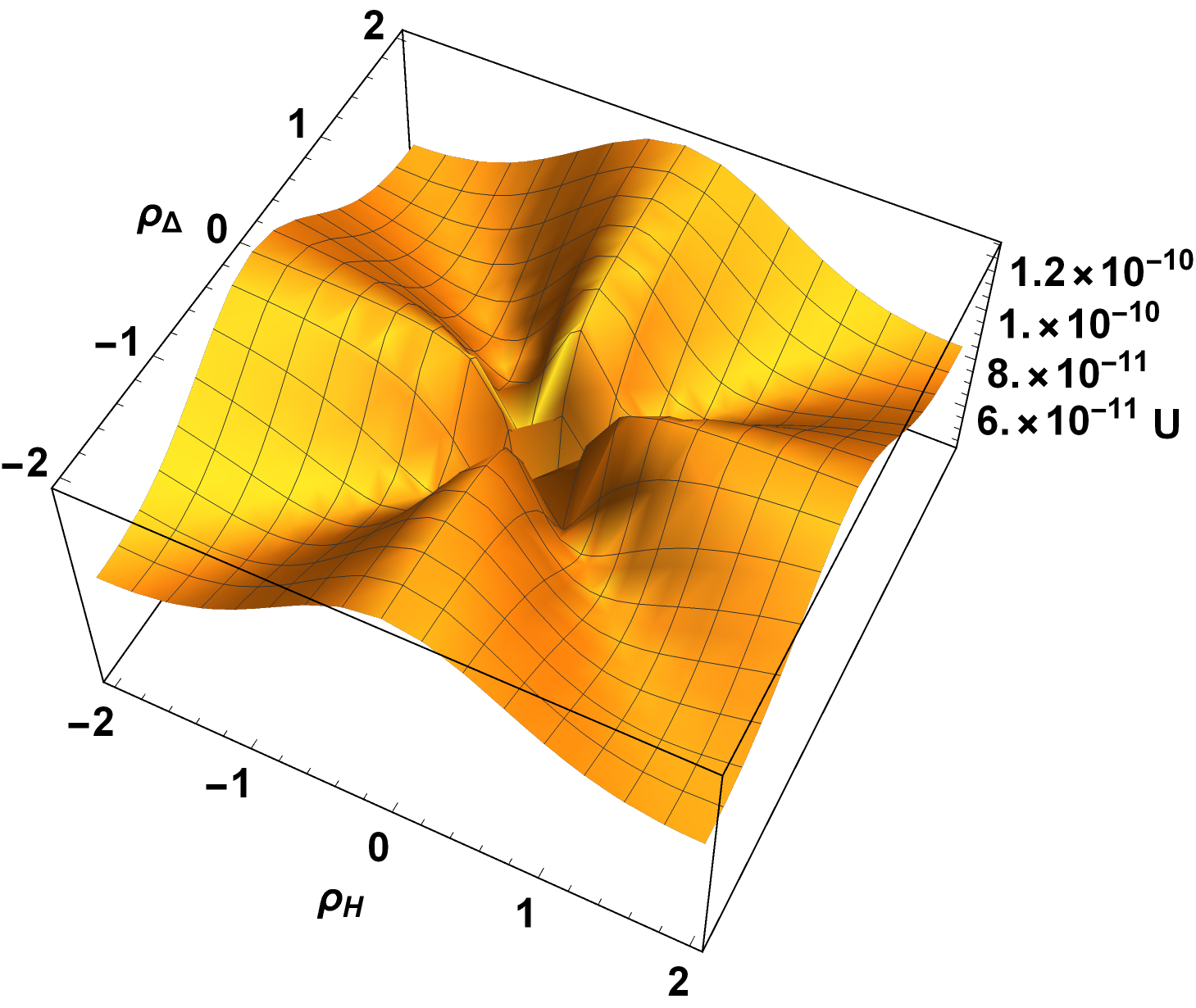}}
    \subfigure[case(2)]{\includegraphics[width=0.48\textwidth]{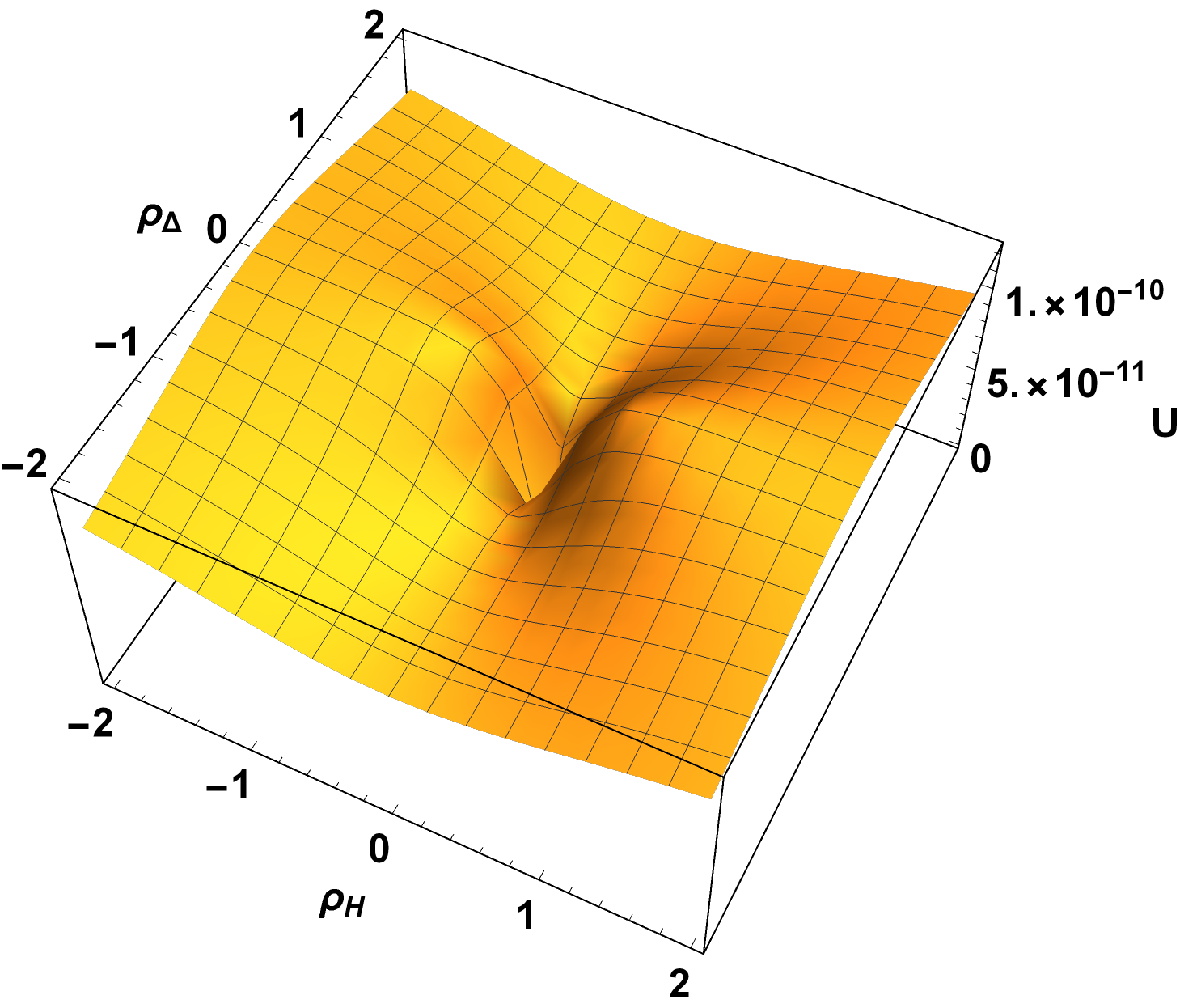}}
    \\
    \subfigure[case(3)]{\includegraphics[width=0.48\textwidth]{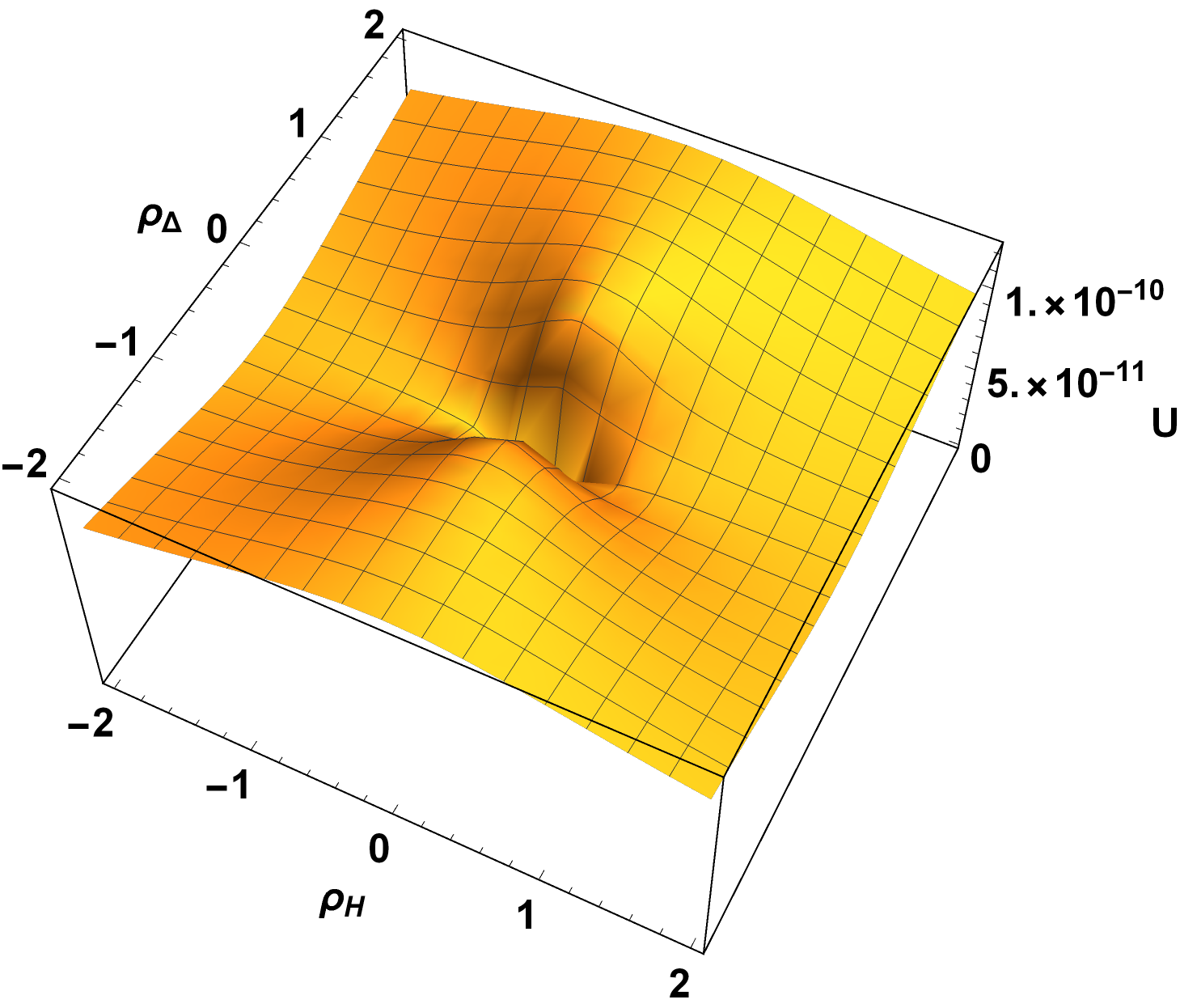}}
    \subfigure[case(4)]{\includegraphics[width=0.48\textwidth]{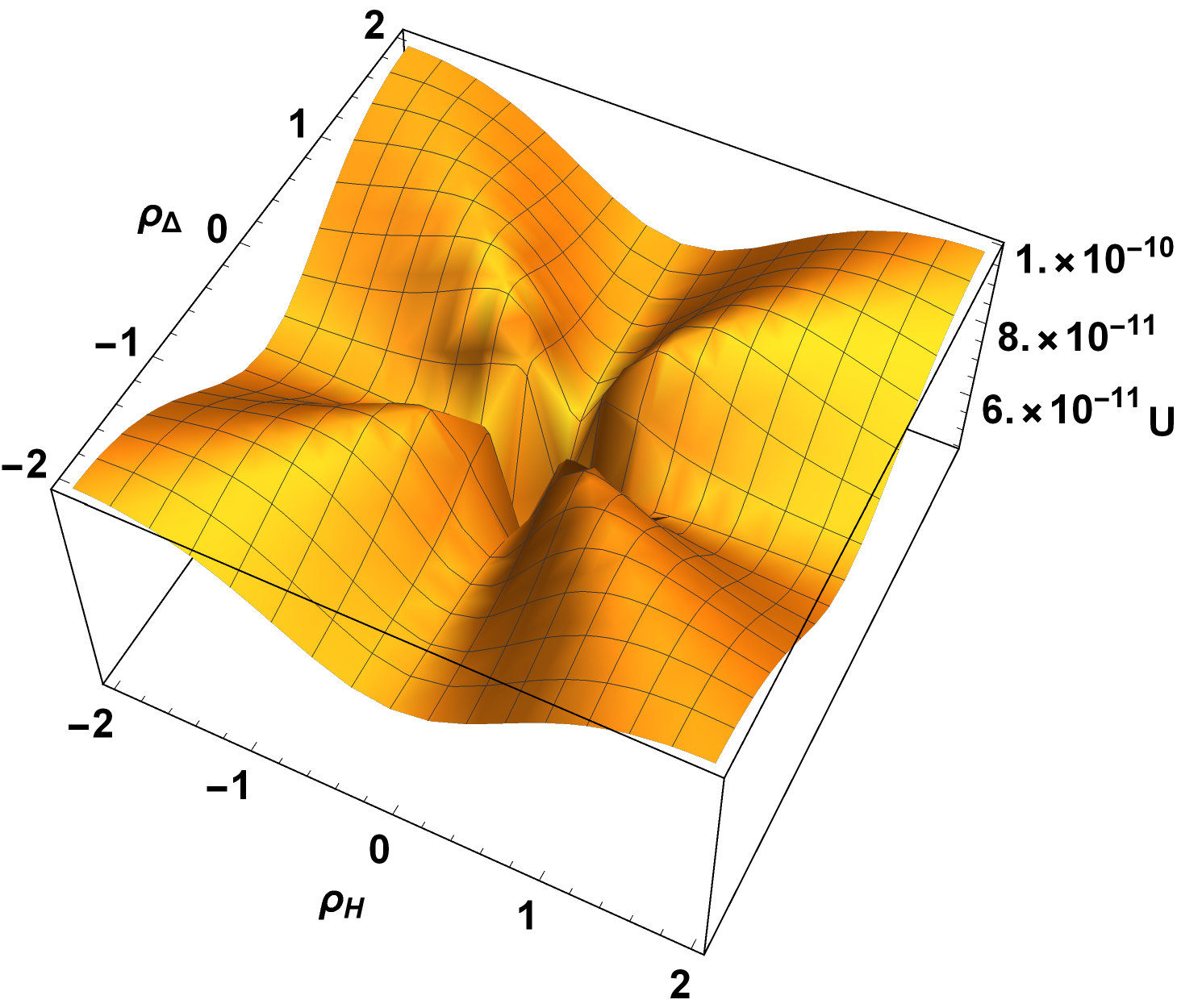}}
    \caption{\label{fig:potential}The potential $U(\rho_H,\rho_\Delta)$ with different minima.}
\end{figure}

Taking the large field limit, the quartic terms in the potential dominate, and the scalar potential in the Einstein frame reduces to 
\begin{equation}
    U(\rho_H,\rho_\Delta)=\frac{V}{\Omega^4}=\frac{\lambda_H\kappa^4+\lambda_{H\Delta}\kappa^2+\lambda_\Delta}{4(\xi_H\kappa^2+\xi_\Delta)^2}~.
\end{equation}
It can be shown that the mass of $\kappa^\prime$ is of order $1/\sqrt{\xi}$, always larger than the Hubble parameter during inflation which is of order $1/\xi$. Therefore, the field $\kappa$ can be integrated out, and the inflation dynamics is determined by $\chi$.

As pointed by~\cite{Lebedev:2011aq},  the potential $U(\rho_H,\rho_\Delta)$ has the following minima,
\begin{eqnarray}
    \label{con:minima}
    &&(1)~2\lambda_H\xi_\Delta-\lambda_{H\Delta}\xi_H>0,~2\lambda_\Delta\xi_H-\lambda_{H\Delta}\xi_\Delta>0,~\kappa=\sqrt{\frac{2\lambda_\Delta\xi_H-\lambda_{H\Delta}\xi_\Delta}{2\lambda_H\xi_\Delta-\lambda_{H\Delta}\xi_H}}~,\\
    &&(2)~2\lambda_H\xi_\Delta-\lambda_{H\Delta}\xi_H>0,~2\lambda_\Delta\xi_H-\lambda_{H\Delta}\xi_\Delta<0,~\kappa=0~,\\
    &&(3)~2\lambda_H\xi_\Delta-\lambda_{H\Delta}\xi_H<0,~2\lambda_\Delta\xi_H-\lambda_{H\Delta}\xi_\Delta>0,~\kappa=\infty~,\\
    &&(4)~2\lambda_H\xi_\Delta-\lambda_{H\Delta}\xi_H<0,~2\lambda_\Delta\xi_H-\lambda_{H\Delta}\xi_\Delta<0,~\kappa=0,\infty~.
\end{eqnarray}
As shown in Fig.~\ref{fig:potential}, in case (1) the minimum is in the direction of the mixing of $h$ and $\Delta^0$, while in other cases the valley extends in the direction of $h$ or $\Delta^0$. If the inflaton evolves along the valley, as will be discussed in Sec. \ref{sec:valley}, the inflation dynamics is essentially equivalent to the single-field  inflation. In this situation, successful leptogenesis can only occur in case (1). However, as will be discussed in Sec. \ref{sec:ridge}, if a local minimum exhibits in the ridge in case (2), then successful leptogenesis can also be achieved. Moreover, in this case the single-field  inflation approximation is invalid, and some interesting features of multifield inflation arise.

\section{The Affleck-Dine Leptogenesis}
\label{sec:leptogenesis}
In the Affleck-Dine mechanism, the baryon asymmetry is generated through a nonzero angular motion of phases of the scalar fields carrying $U(1)_B$ charge. In our model the lepton asymmetry is generated analogously by the angular rotation of the phase of $\Delta^0$ when it evolves along the inflation trajectory. Thus, to analyse leptogenesis, we should keep track of the evolution of the phases by solving the Klein-Gordon equation Eq.~\ref{eq:back_evolution} with $\varphi^I=(\rho_H,\eta,\rho_\Delta,\theta)$. The generated lepton number density during this process can be calculated through the Noether current,
\begin{equation}
    n_L=j_L^0=\frac{Q_L \rho^2_\Delta \dot{\theta}}{1+\xi_H\rho_H^2+\xi_\Delta\rho_\Delta^2}.
\end{equation}
After reheating~\cite{Garcia-Bellido:2008ycs, Bezrukov:2008ut,Ema:2016dny, DeCross:2015uza, DeCross:2016cbs, DeCross:2016fdz, He:2018mgb,He:2020ivk,He:2020qcb, Sfakianakis:2018lzf}, the generated lepton number density will be present as SM particles. Then, through the equilibrium electroweak sphaleron process before electroweak symmetry breaking, part of lepton asymmetry is converted into baryon asymmetry, $n_B = -\frac{28}{79}n_L$~ \cite{Klinkhamer:1984di,Kuzmin:1985mm,Trodden:1998ym,Sugamoto:1982cn}. As pointed by~~\cite{Barrie:2021mwi,Barrie:2022cub}, a lepton number density $n_L\sim 10^{-16} M_P^{-3}$ at the end of the inflation could just explain the baryon asymmetry of our universe.

To have non-trivial motion of $\theta$ during inflation, we need the presence of $U(1)_L$ breaking term during inflation. This condition requires the motion of the inflation to occur along a path where both $h$ and $\Delta^0$ are non-vanishing. One obvious choice is that if the conditions in case (1) are satisfied, the inflation will follow a direction with fixed $\kappa$ and the $U(1)_L$ breaking condition is satisfied. This scenario is pointed out by \cite{Barrie:2021mwi,Barrie:2022cub}  and we denote it as the ``canonical scenario" of type-II seesaw leptogenesis. Note that in the previous studies \cite{Barrie:2021mwi,Barrie:2022cub}, the estimation of the lepton asymmetry is calculated via the single-field  approximation. 
Instead, in this work we adopt a full calculation without any approximation, and we will show that the single-field  approximation indeed work very well, supporting the result of \cite{Barrie:2021mwi,Barrie:2022cub}.  For the other three cases,  it seems that the inflation will occur in a direction where either $h$ or $\Delta^0$ is nearly zero and thus the motion of $\theta$ is strongly suppressed because the $U(1)_L$ breaking term is too small. However, we will show that there exist another scenario that the lepton asymmetry can be generated if the inflaton evolves along the ridge. In this scenario, the single field approximation breaks down and a sizable non-Gaussianity could be generated. 

\subsection{Along the valley}
\label{sec:valley}
As discussed in the preceding section, the field $\kappa$ can be effectively integrated out, and the minima of the potential extend in some specific directions with a fixed $\kappa$. Then the inflaton evolves along the valley and a single-field approximation can be used during the inflation.  In order to achieve successful leptogenesis, the valley should extend in the direction of the mixing of $h^0$ and $\Delta^0$, as in case (1). In this work, instead of using the single-field  approximation, we present a full numerical calculation to solve the evolution of the fields. 

\begin{table}[htbp]
  \centering
  \caption{A benchmark point in the parameter region satisfying case (1).}
  \vspace{.2cm}
  \label{tab:para_valley}
  \begin{tabular}{|c|c|c|c|c|c|c|c|}
    \hline
    $\xi_H$ & $\xi_\Delta$ & $\lambda_H$ & $\lambda_\Delta$ & $\lambda_{H\Delta}$ & $\mu$ & $\lambda_5$ & $\lambda_5^\prime$\\
    \hline
    $300$ & $300$ & $4\times10^{-5}$ & $6\times 10^{-5}$ & $4\times 10^{-5}$ & $0$ & $0$ & $-10^{-9}$\\
    \hline
  \end{tabular}
\end{table}

\begin{figure}[htbp]
    \centering
    \subfigure{\includegraphics[width=0.47\textwidth]{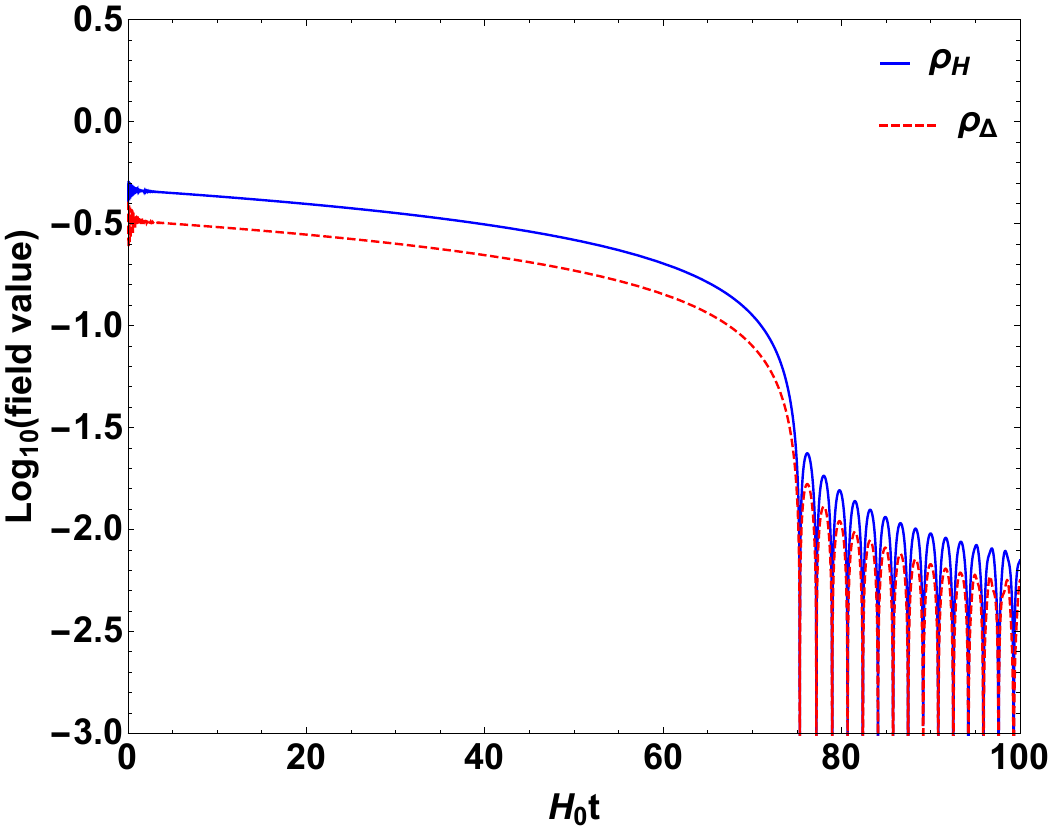}}
    \hspace{.3cm}
    \subfigure{\includegraphics[width=0.45\textwidth]{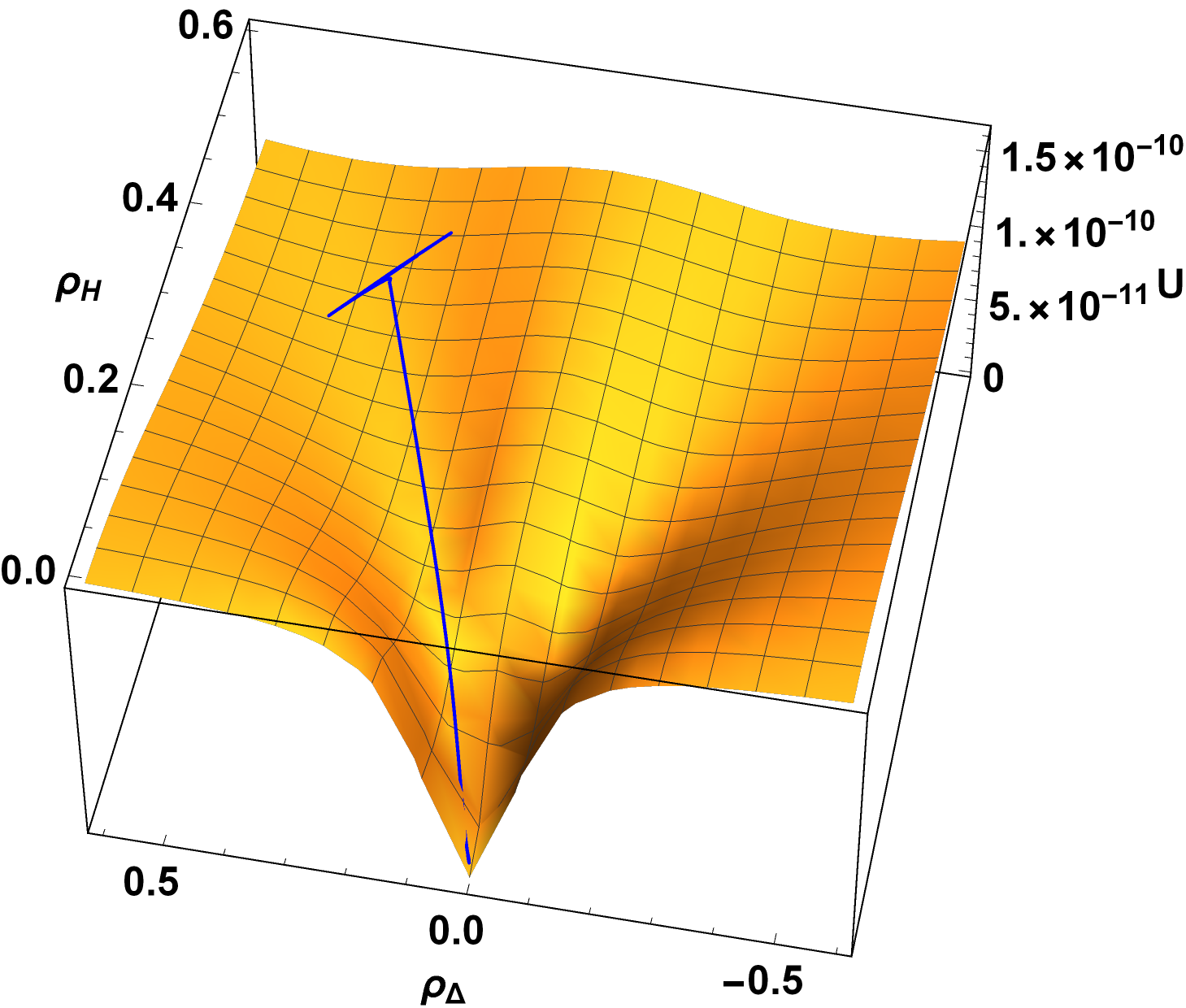}}
    \caption{\label{fig:field_valley}The evolution of the inflaton along the valley.}
\end{figure}

In Table.~\ref{tab:para_valley} we show one benchmark point in the parameter space satisfying the conditions in case (1). With the initial conditions chosen as $\rho_H=0.4, \rho_\Delta=0.4, \eta=0, \theta=0.1, \dot{\rho}_H=\dot{\rho}_\Delta=\dot{\eta}=\dot{\theta}=0$, the evolution of the inflaton is solved and depicted in Fig.~\ref{fig:field_valley}. We have parameterized the evolution with respect to $H_0 t$, with $H_0$ defined as the Hubble constant when the inflation starts. The inflation ends at $H_0 t=74.44$, leading to the number of efoldings of expansion $N=70.22$.

\begin{figure}[htbp]
    \centering
    \subfigure{\includegraphics[width=0.55\textwidth]{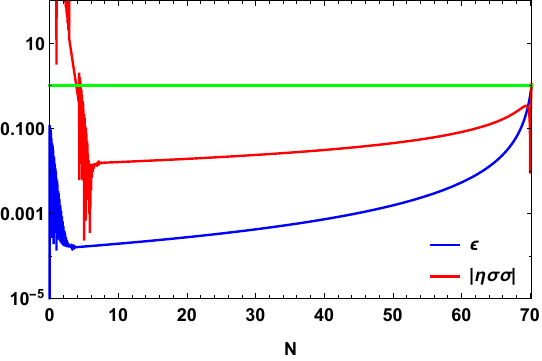}}
    \caption{\label{fig:SR_valley}The slow-roll parameters when the inflaton evolves along the valley.}
\end{figure}

The slow-roll parameters and the turn rate are depicted in Fig.~\ref{fig:SR_valley}. It is obvious that, after a short period of rolling down to the valley, the slow-roll parameters and the turn rate quickly approach the slow-roll slow-turn limit $\epsilon, |\eta_{\sigma\sigma}|, |\omega^I|\ll 1$. Then, as the inflaton evolves along the valley, the single-field approximation is valid and the inflation is effectively standard single-field inflation.

The spectral index of the power spectrum at the horizon crossing $N_\star=N_{end}-60$ is also calculated, leading to $n_s(N_\star)=0.967$ which is in excellent agreement with current CMB observations. 

\begin{figure}[htbp]
    \centering
    \includegraphics[width=0.55\textwidth]{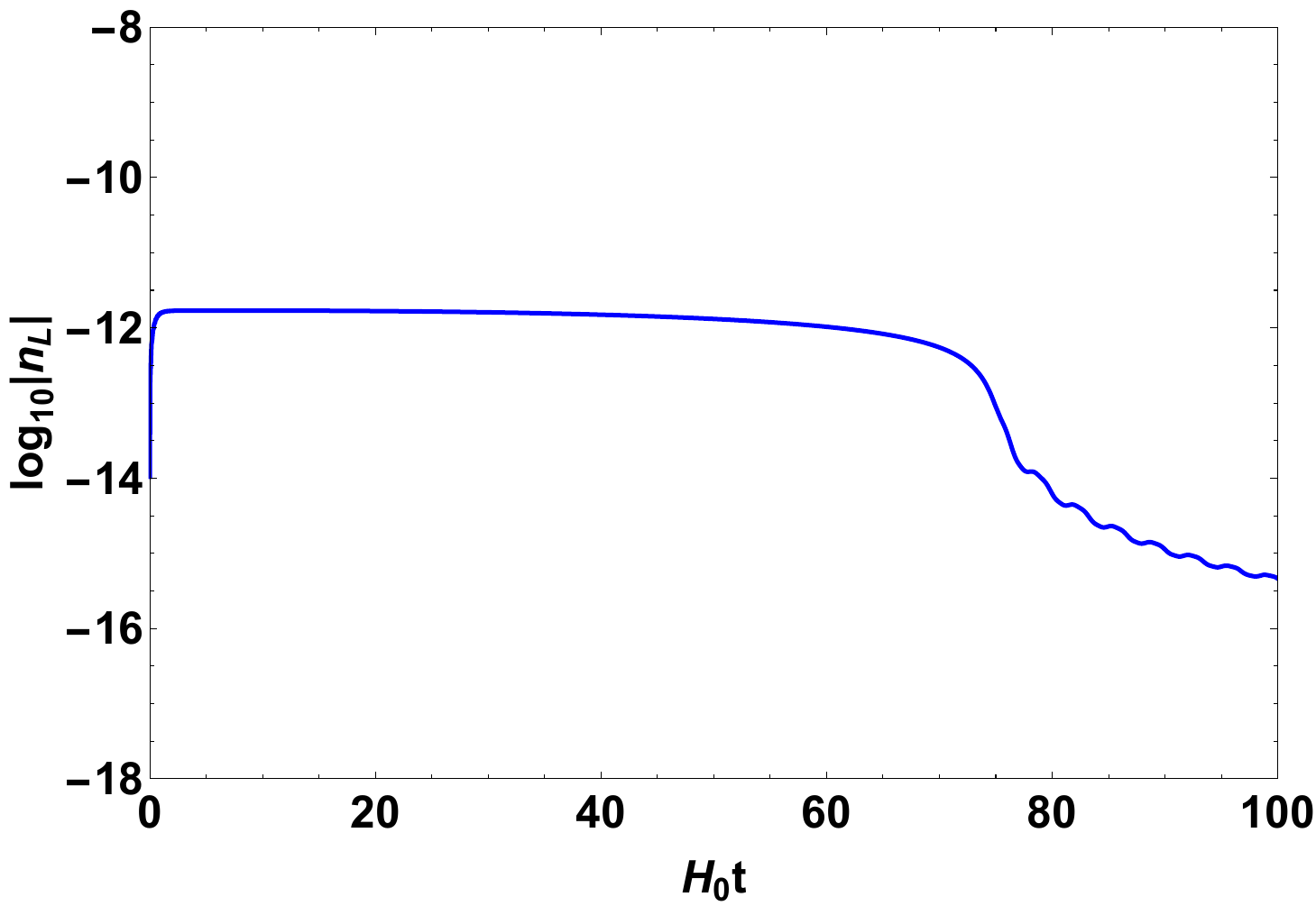}
    \caption{\label{fig:lepton_density_valley}The generated lepton number density during inflation along the valley.}
\end{figure}

Since the evolution of inflaton is solved, the lepton number generated during inflation can be determined, which is depicted in Fig.~\ref{fig:lepton_density_valley}. We see that at this benchmark point an adequate lepton asymmetry is generated. By simply setting $\lambda^\prime_5=-10^{-12}$, we can obtain $|n_L(t_{end})|\sim 10^{-16} M_P^{-3}$, which is required to explain the baryon asymmetry observed today. 

\subsection{Along the ridge}
\label{sec:ridge}
As demonstrated above,  typically the inflaton rapidly descends to the valley, evolving as a single field. However, exceptions exist. If a local minimum exists along the ridge and the inflaton initially resides there, it may traverse the ridge before ultimately transitioning to the valley. Should this transition occur in the late stages, the CMB data remains explicable, and successful leptogenesis is attainable. In such cases, the single-field approximation becomes invalid.


Here we take the case (2) as an example, where the ridge extends in the direction of $\rho_H$. A local minimum on the ridge requires the potential
\begin{equation}
\label{con:local_minima}
    \left.\frac{\partial^2 U}{\partial \rho_\Delta^2}\right|_{\rho_\Delta=0}=\frac{\rho_H^2}{2(1+\xi_H \rho_H^2)^3}\left[\lambda_{H\Delta} + (\lambda_{H\Delta} \xi_H-2\lambda_H \xi_\Delta)\rho_H^2\right]>0~.
\end{equation}

\begin{table}[htbp]
  \centering
  \caption{A benchmark point in the parameter region satisfying case (2) and Eq.~\ref{con:local_minima}.} 
  \vspace{.3cm} 
  \label{tab:para_ridge}
  \begin{tabular}{|c|c|c|c|c|c|c|c|}
    \hline
    $\xi_H$ & $\xi_\Delta$ & $\lambda_H$ & $\lambda_\Delta$ & $\lambda_{H\Delta}$ & $\mu$ & $\lambda_5$ & $\lambda_5^\prime$\\
    \hline
    $100$ & $100.02$ & $5\times10^{-6}$ & $3.5\times 10^{-6}$ & $1\times 10^{-5}$ & $0$ & $0$ & $-10^{-10}$\\
    \hline
  \end{tabular}
\end{table}

\begin{figure}[htbp]
    \centering
    \subfigure{\includegraphics[width=0.47\textwidth]{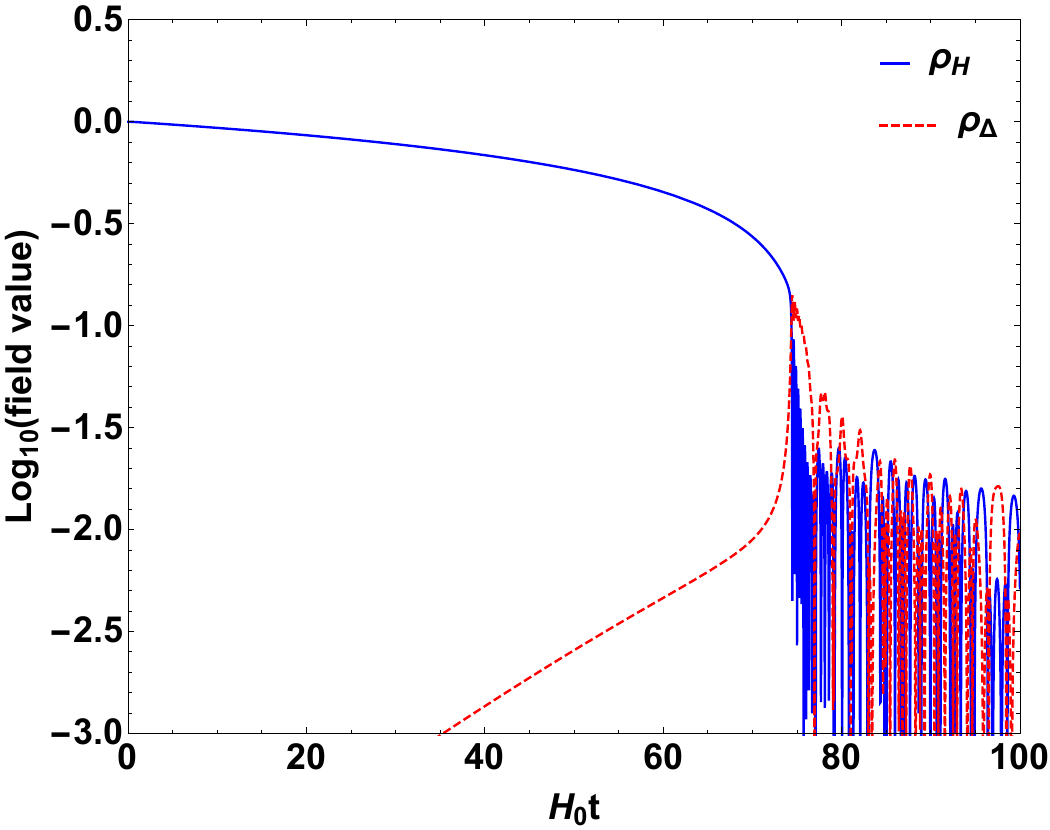}}
    \hspace{.3cm} 
    \subfigure{\includegraphics[width=0.45\textwidth]{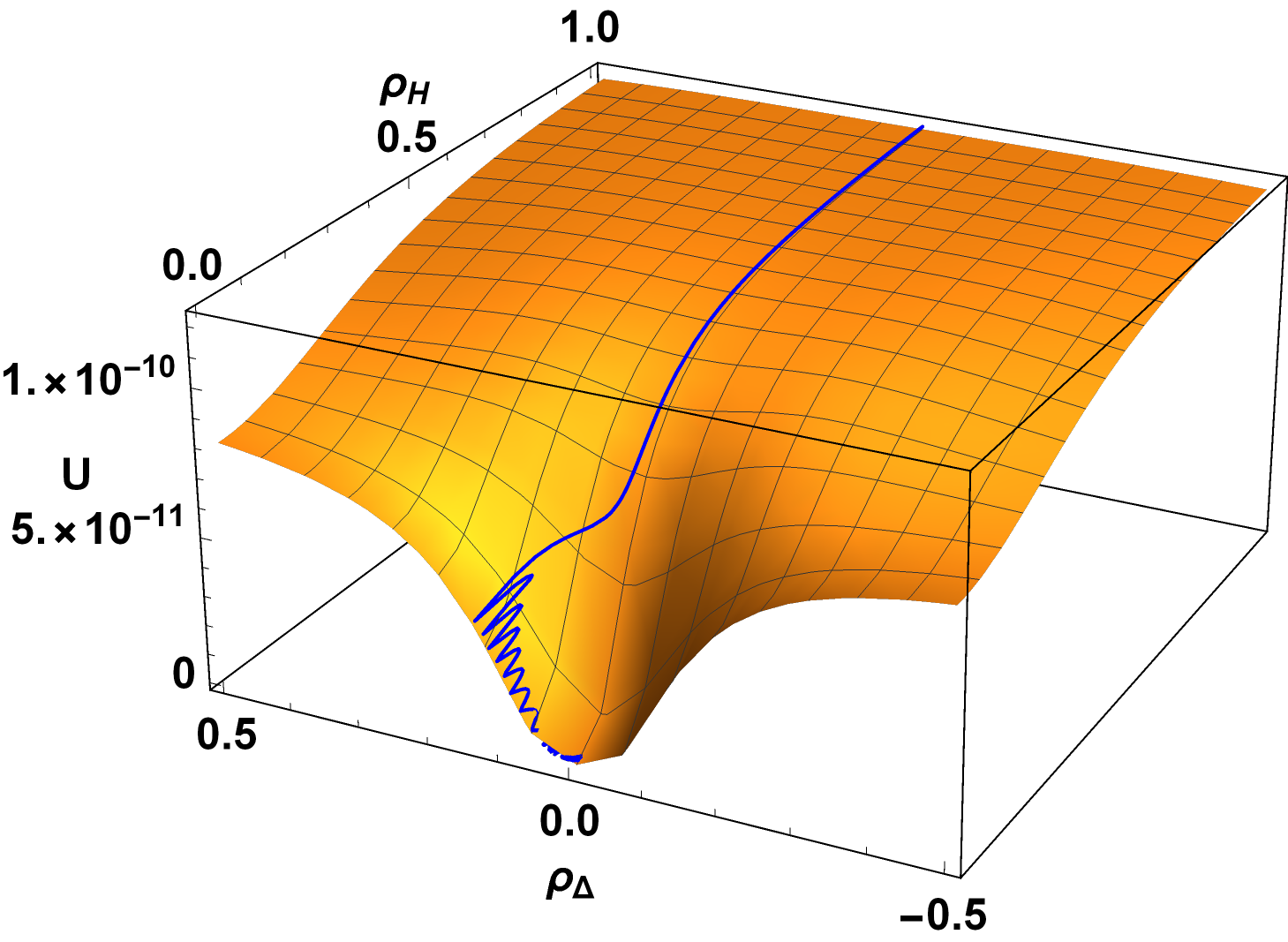}}
    \caption{\label{fig:field_ridge}The evolution of the inflaton along the ridge.}
\end{figure}

In Table ~\ref{tab:para_ridge} we present a benchmark point in the parameter region satisfying the conditions in case (2) and Eq.~\ref{con:local_minima}. With the initial condition chosen as $\rho_H=1, \rho_\Delta=8.7\times 10^{-5}, \eta=0, \theta=0.1, \dot{\rho}_H=\dot{\rho}_\Delta=\dot{\eta}=\dot{\theta}=0$, the evolution of the inflaton is solved and depicted in Fig.~\ref{fig:field_ridge}. The inflation ends at $H_0 t=75.95$, leading to the number of efoldings of expansion $N=73.31$. The spectral index in this case, $n_s(N_\star)=0.967$, is also in excellent agreement with current CMB observations.

\begin{figure}[htbp]
    \centering
    \subfigure{\includegraphics[width=0.55\textwidth]{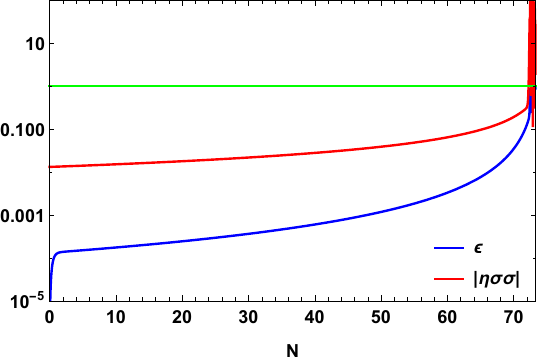}}
    \caption{\label{fig:SR_ridge}The slow-roll parameters when the inflaton evolves along the valley.}
\end{figure}

\begin{figure}[htbp]
    \centering
    \includegraphics[width=0.55\textwidth]{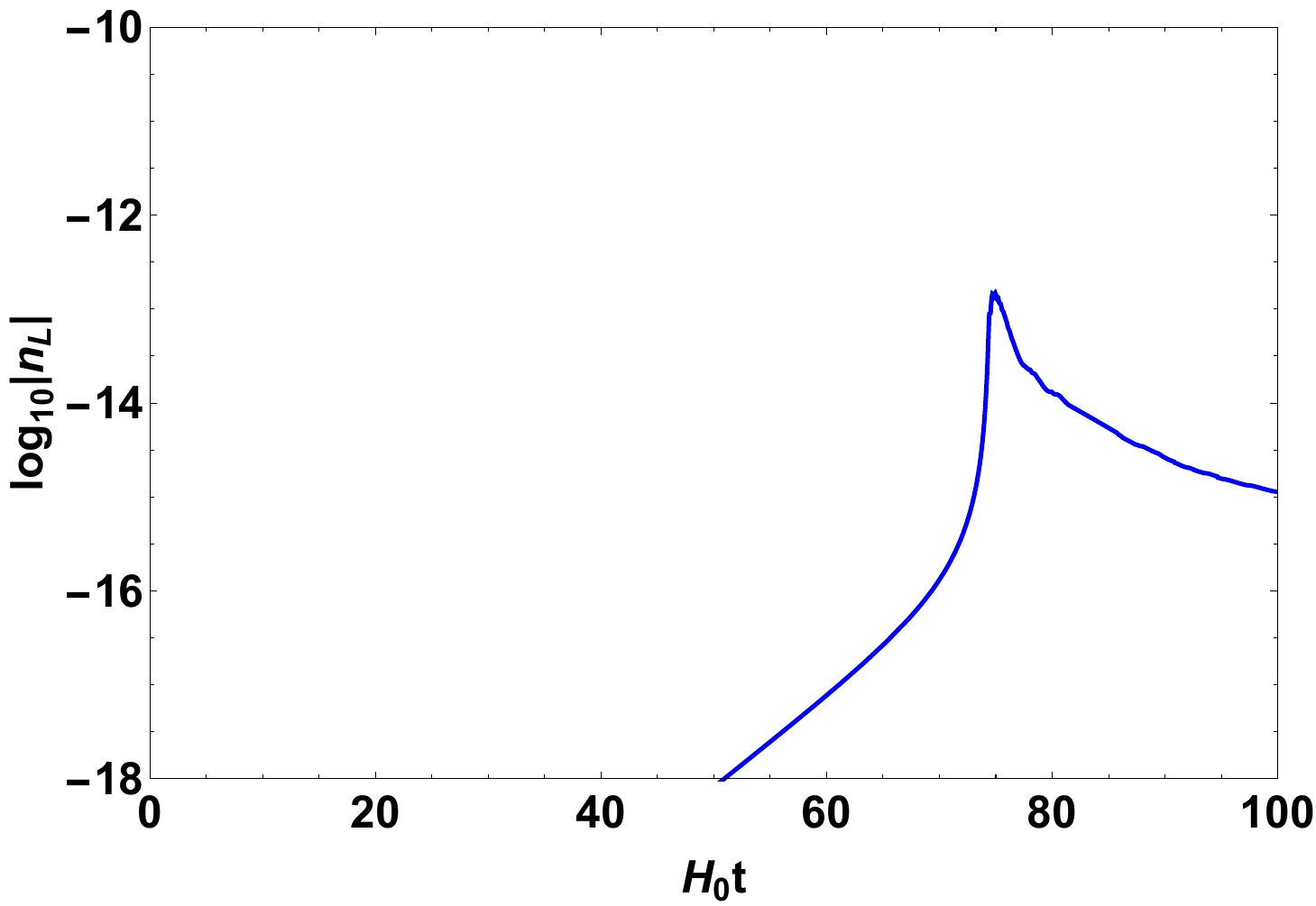}
    \caption{\label{fig:lepton_density_ridge}The generated lepton number density during inflation along the ridge.}
\end{figure}

The slow-roll parameters and the turn rate are depicted in Fig.~\ref{fig:SR_ridge}. One can observe that when the inflaton evolves along the ridge, the slow-roll slow-turn limit $\epsilon, |\eta_{\sigma\sigma}|, |\omega^I|\ll 1$ is satisfied. However, when the inflaton rolls down the ridge, the deviation from the single-field inflation leads to something interesting. As depicted in Fig.~\ref{fig:lepton_density_ridge}, when the inflaton rolls off, the rapid turn in the field space leads to a rapid motion of $\theta$, and thus a large amount of lepton number is generated at this time. To match the observed baryon asymmetry today, we can simply set $\lambda^\prime_5=-3\times10^{-13}$ in this case.

\begin{figure}[htbp]
    \centering
    \subfigure{\includegraphics[width=0.48\textwidth]{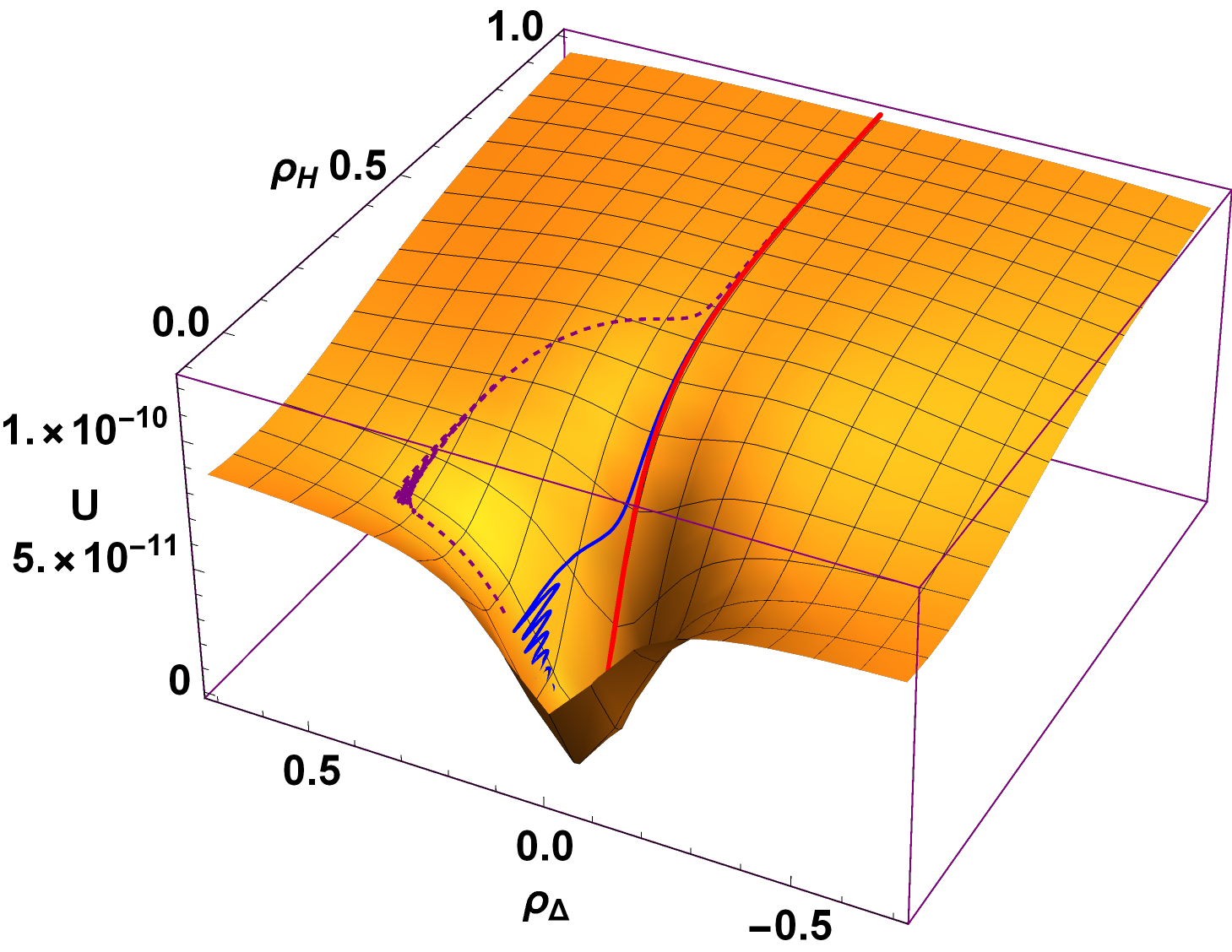}}
    \subfigure{\includegraphics[width=0.48\textwidth]{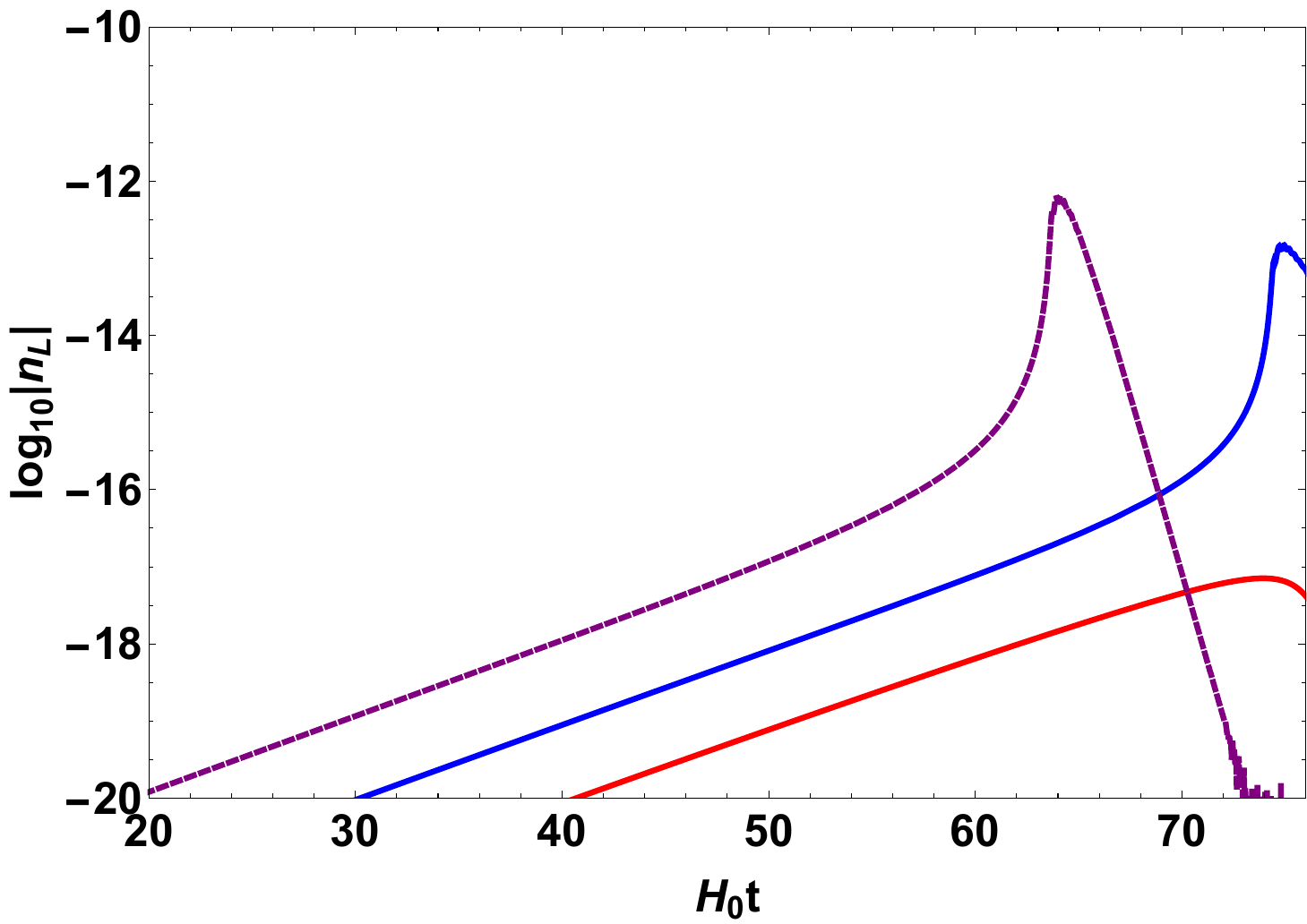}} 
    \caption{\label{fig:ridge_vari}The type-II seesaw leptogenesis along the ridge with different initial conditions. The red, blue, and purple lines correspond to initial field value $\rho_\Delta=4\times 10^{-5},~8.7\times 10^{-5},~2\times 10^{-4}$ respectively.}
\end{figure}

For the aforementioned benchmark point, we deliberately selected a very small value for $\rho_\Delta$ to ensure that the inflaton remains on the ridge for an extended duration. However, varying this parameter can significantly alter the trajectory of inflation. In Fig.~\ref{fig:ridge_vari}, we depict the inflationary paths for different initial conditions of $\rho_\Delta$, specifically $4\times 10^{-5}$, $8.7\times 10^{-5}$, and $2\times 10^{-4}$. The results indicate that deviating the initial $\rho_\Delta$ significantly from the local minima causes the inflaton to descend to the valley much earlier. Subsequently, it follows the valley where $\rho_h=0$, resulting in a substantial suppression of lepton number generation. Simultaneously, the previously generated lepton asymmetry is diluted during the late-stage inflation. This combined effect leads to an exceedingly small lepton asymmetry in such instances. The right panel of Fig.~\ref{fig:ridge_vari} illustrates this, with the purple line indicating a rapid decrease in the generated lepton number during the late stage of inflation.

Conversely, choosing the initial $\rho_\Delta$ too close to the local minima results in a continuous stay at this point without any turning. In such cases, the generated lepton asymmetry remains consistently small, as depicted by the red line on the right panel of Fig.~\ref{fig:ridge_vari}. In this case, a larger $\lambda_5$ seems needed to get the correct value of baryon asymmetry in our universe.


\section{Non-Gaussianity}
\label{sec:nongauss}
In multifield inflation, an important prediction is the primordial non-Gaussianity of cosmological perturbations~\cite{Malik:2008im, Wands:2007bd, Bartolo:2004if, Chen:2010xka, Byrnes:2010em, Komatsu:2009kd, Qiu:2010dk, Bernardeau:2002jy, Bernardeau:2002jf,Langlois:2008mn, Seery:2005gb, Yokoyama:2007uu, Yokoyama:2007dw, Byrnes:2008wi, Elliston:2011dr, Seery:2012vj, Peterson:2010np, Peterson:2010mv, Peterson:2011yt, Gong:2011cd, Mazumdar:2012jj}. The non-Gaussianity leads to nonzero three-point or higher order correlation functions. In the following, we calculate the primordial bispectra produced by the inflation processes discussed above. 

In order to handle field perturbations in the models with nontrivial field-space manifolds, we adopt the covariant formalism~\cite{Gong:2011uw,Kaiser:2012ak,Elliston_2012}. The field perturbation $\delta \phi^I=\phi^I-\varphi^I$ does not transform covariantly. However, after parametrizing the geodesic connecting $\varphi^I$ and $\phi^I$ by $\lambda$ such that $\phi^I(\lambda=0)=\varphi^I$ and $\phi^I(\lambda=1)=\phi^I$, $\delta\phi^I$ can be expanded as
\begin{equation}
 \delta \phi^{I}=\mathcal{Q}^{I}-\frac{1}{2} \Gamma_{J K}^{I} \mathcal{Q}^{J} \mathcal{Q}^{K}+\frac{1}{6}\left(\Gamma_{L M}^{I} \Gamma_{J K}^{M}-\Gamma_{J K ; L}^{I}\right) \mathcal{Q}^{J} \mathcal{Q}^{K} \mathcal{Q}^{L}+\cdots,
\end{equation}
where $\mathcal{Q}^I=\left.\frac{d\phi^I}{d\lambda}\right|_{\lambda=0}$ is the tangent vector. We also adopt the gauge invariant Mukhanov-Sasaki variable $Q^I=\mathcal{Q}^I+\psi\frac{\dot{\varphi}^I}{H}$, where $\psi$ is the scalar component of spatial metric perturbation. In the spatially flat gauge up to first order, $Q^I=\mathcal{Q}^I=\delta\phi^I$.

The quantity of interest is the bispectrum of the curvature perturbation $\zeta$, parametrized as 
\begin{equation}
\label{eq:bispec_zeta}
\langle\zeta(\boldsymbol{k}_1)\zeta(\boldsymbol{k}_2)\zeta(\boldsymbol{k}_3)\rangle=S(k_1,k_2,k_3)\frac{1}{(k_1k_2k_3)^2}\mathcal{P}_{\zeta}^2(2\pi)^7\delta^3(\boldsymbol{k}_1+\boldsymbol{k}_2+\boldsymbol{k}_3),
\end{equation}
where $\mathcal{P}_{\zeta}$ is the power spectrum of $\zeta$ and the $S(k_1,k_2,k_3)$ is the shape function.

Adopting the $\delta N$ formalism~\cite{Sasaki:1995aw, Lyth:2004gb, Lee:2005bb}, $\zeta$ on super-Hubble scales can be expanded as a function of the field perturbations on the initial flat hypersurface,
\begin{equation}
    \zeta(x)=N_{,I}Q^I(x)+\frac{1}{2}N_{,IJ}Q^I(x) Q^J(x)+\cdots,
\end{equation}
where $N_{,I}=\left.\mathcal{D}_I N\right|_{\varphi^I_\star},~N_{,IJ}=\left.\mathcal{D}_I\mathcal{D}_J N\right|_{\varphi^I_\star,\varphi^J_\star}$ and $N$ is the e-fold number between the initial flat hypersurface and the final uniform energy density hypersurface. Then the power spectrum and the bispectrum can be expanded as
\begin{equation}
\left\langle\zeta\left(\boldsymbol{k}_{1}\right) \zeta\left(\boldsymbol{k}_{2}\right)\right\rangle=(2 \pi)^{5} \frac{\mathcal{P}_{\zeta}}{2 k_{1}^{3}} \delta^{3}\left(\boldsymbol{k}_{1}+\boldsymbol{k}_{2}\right),\quad \mathcal{P}_\zeta=\left(\frac{H_\star}{2\pi}\right)^2N_{,I}N^{,I},
\end{equation}
\begin{equation}
\begin{aligned}
\langle\zeta(\boldsymbol{k}_1)\zeta(\boldsymbol{k}_2)\zeta(\boldsymbol{k}_3)\rangle=&N_{,I}N_{,J}N_{,K}\langle Q^I(\boldsymbol{k}_1)Q^J(\boldsymbol{k}_2)Q^K(\boldsymbol{k}_3)\rangle_\star\\
&+\frac{1}{2}N_{,IJ}N_{,K}N_{,L}\int\frac{d^3q}{(2\pi)^3}\langle Q^I(\boldsymbol{k}_1-\boldsymbol{q})Q^K(\boldsymbol{k}_2)\rangle_\star\langle Q^J(\boldsymbol{q})Q^L(\boldsymbol{k}_3)\rangle_\star\\
&+\mathrm{cyclic~permutations}.
\end{aligned}
\end{equation}

We can see that there are two contributions to the bispectrum of $\zeta$. One results from the linear transfer of the intrinsic bispectra of the field perturbations at horizon crossing, corresponding to the first term at the right side. The other results from the non-linear relation between the Gaussian field perturbations $\zeta$, corresponding to the other terms. The in-in formalism is needed to calculate $\langle Q^I(\boldsymbol{k}_1)Q^J(\boldsymbol{k}_2)Q^K(\boldsymbol{k}_3)\rangle_\star$. Luckily, it was shown that the first contribution remains considerably smaller than the latter contribution for the family of models of interest~\cite{Kaiser:2012ak}, thus we can reasonably assume that at horizon crossing, the perturbation is Gaussian. Then the bispectrum of $\zeta$ reduces to
\begin{equation}
    \left\langle\zeta\left(\mathbf{k}_{1}\right) \zeta\left(\mathbf{k}_{2}\right) \zeta\left(\mathbf{k}_{3}\right)\right\rangle=N_{,IJ} N^{,I} N^{,J} \frac{H_{*}^{4}}{4}\left(\frac{1}{k_{1}^{3} k_{2}^{3}}+\frac{1}{k_{2}^{3} k_{3}^{3}}+\frac{1}{k_{3}^{3} k_{1}^{3}}\right)(2 \pi)^{3} \delta^{3}\left(\boldsymbol{k}_{1}+\boldsymbol{k}_{2}+\boldsymbol{k}_{3}\right) .
\end{equation}

According to Eq. ~\ref{eq:bispec_zeta}, the shape function is
\begin{equation}
    S(k_1,k_2,k_3)=\frac{1}{4}\frac{N_{,IJ} N^{,I} N^{,J}}{(N_{,K}N^{,K})^2}\left(\frac{k_1^2}{k_2k_3}+\mathrm{cyclic~permutations}\right).
\end{equation}
Shape function of this type is called local shape, which peaks at the squeezed triangle limit. Conventionally, the amplitude of bispectrum is denoted as $f_{NL}$ by matching to the shape function,
\begin{equation}
    S\left(k_{1}, k_{2}, k_{3}\right) \underset{\text { limit }}{\stackrel{k_{1}=k_{2}=k_{3}}{\longrightarrow}} \frac{9}{10} f_{N L} \text {. }
\end{equation}
In this case, $f_{NL}$ is given by
\begin{equation}
	f_{N L}=-\frac{5}{6} \frac{N^{, I} N^{, J} \mathcal{D}_{I} \mathcal{D}_{J} N}{\left(N_{, K} N^{, K}\right)^{2}}.
\end{equation}

\begin{figure}[htbp]
    \centering
    \includegraphics[width=0.8\textwidth]{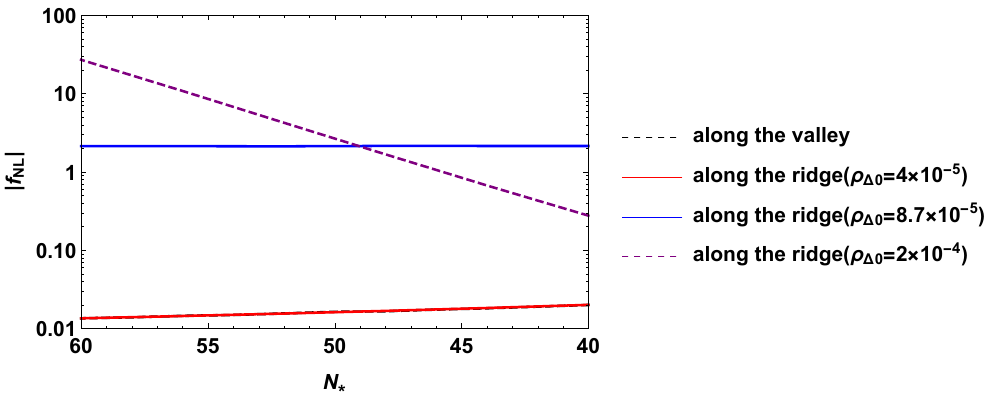}
    \caption{\label{fig:nongaussian}The non-Gaussianity produced in the inflation processes along the valley and the ridge.}
\end{figure}

In general multifield models, it is difficult to find the analytical form of $\delta N$ formalism, thus we use finite difference method to perform the calculation numerically. For example, $N_{,\rho_H}$ can be obtained by
\begin{equation}
	N_{, \rho_H}=\frac{N(\rho_H+\Delta \rho_H, \rho_\Delta)-N(\rho_H-\Delta \rho_H, \rho_\Delta)}{2 \Delta \rho_H},
\end{equation}
where $N(\rho_H, \rho_\Delta)$ is the number of e-folds from horizon-crossing to the end of inflation, and $\rho_H, \rho_\Delta$ are the initial values at horizon crossing. We give a perturbation $\Delta\phi^I$ to the initial value, and numerically solve the classical field equation of motion to obtain $N(\phi^I+\Delta\phi^I)$. Then, after tuning $\Delta\rho_H$ or $\Delta\rho_\Delta$ to sufficiently small, the numerical result can converge, as depicted in Fig.~\ref{fig:nongaussian}. 

We see that when the inflaton evolves along the valley, the generated non-Gaussianity is negligible. However, when the inflaton evolves along the ridge, the generated non-Gaussianity can be as large as $|f_{NL}|\sim 2.1$ for fiducial $k_\star$ which first crossed the horizon $N_\star=60$ e-folds before inflation ends. Given current limit $f_{NL}$ is $-0.9 \pm 5.1$ for local shape~\cite{Planck:2019kim}, such a large non-Gaussianity may be observed and tested in the upcoming CMB experiments.

\section{Conclusions}
\label{sec:conclusion}

Type-II seesaw leptogenesis is a model that simultaneously explains inflation, baryon number asymmetry, and neutrino mass, employing the Affleck-Dine mechanism to generate lepton asymmetry and using the Higgs bosons as the inflaton. Previous studies assumed inflation to occur in a valley of the potential, taking the single-field approximation. In this work, we explored an alternative scenario for such type-II seesaw leptogenesis, where the inflation takes place along a ridge of the potential.  Firstly we conducted a comprehensive numerical calculation in the canonical scenario, where the inflation occurs in a valley, confirming the effectiveness of the single-field approximation. We also introduced a novel scenario wherein the inflation initiates along the potential's ridge and transitions to the valley in the late stages. During the transition, we find the generation of the lepton asymmetry can be enhanced. In this case, the single-field inflation approximation is no longer valid, yet leptogenesis is still successfully achieved. Furthermore, we demonstrated that this scenario can generate a significant non-Gaussianity signature, offering testable predictions for future experiments.

\addcontentsline{toc}{section}{Acknowledgments}
\section*{Acknowledgements}
C. H. is supported by the Guangzhou Basic and Applied Basic Research Foundation under Grant No. 202102020885, the Sun Yat-Sen University Science Foundation and the Fundamental Research Funds for the Central Universities under Grant No. 22qntd3007. This work was also supported by the National Natural Science Foundation of China (NSFC) under grant Nos. 11821505, 12075300 and 1233500, and by Peng-Huan-Wu Theoretical Physics Innovation Center (12047503).

\addcontentsline{toc}{section}{References}
\bibliographystyle{JHEP}
\bibliography{ms}

\end{document}